# Comment on "Energy Velocity of Diffusing Waves in Strongly Scattering Media"

In a recent letter, Schriemer et al. [1] report on the measurements of the mean free paths (transport and scattering), and the energy and group velocities of a suspension of glass beads immersed in water. By using an effective medium model [2] based on a spectral function approach, they can explain all of their experimental results. However, in order to support their effective medium approach [2], they have criticized and misinterpreted the recently introduced effective medium theory [3,4] based on the principle of energy density homogenization.

It is well known [5] that effective medium theories such as the Coherent Potential Approximation (CPA) calculate the average of a given quantity associated with a random medium from a ficticious effective medium which is determined by a self-consistency requirement. In conventional CPA approaches the effective medium is demanded that the forward-scattering amplitude, f(0), of the local difference between scattering and effective medium vanishes on average. Since such an approach concentrates on the average amplitude, i.e., the ballistic part of the wave, they become highly problematic in the regime of strong scattering, i.e., when the wavelength, $\lambda$, of the incident wave is comparable to the size of the scatterers.

To overcome these problems within an effective medium model, we have explicitly chosen [3,4] the averaged energy density homogeneity as the criterion for determining the effective medium. From this criterion it is clear that the new effective medium theory is not a theory for the average amplitude but for the average (diffuse) intensity. Consequently, the new effective medium theory does not suffer from the shortcomings of the conventional CPA. It been applied to scalar, electromagnetic (EM) [3] and acoustic [4] wave propagation in random media with many successes.

For scalar and EM cases, we obtain pronounced dips for the energy transport velocity $v_E$ as a function of frequency for low volume fractions, $f_s$, of high dielectric scatterers [3]. For larger $f_s$ the dips are smeared out, as observed in several experiments. Schriemer et al. [1] erroneously concluded that our new effective medium approach gives the wrong results for acoustic waves, since their measurements indicate considerable structure in $v_E$ at high $f_s$. However, their setup corresponds to low dielectric scatterers in a high dielectric background. For such a configuration all results of the new effective medium theory have to be re-examined: if we calculate $v_E$ for this inverse geometry, we indeed find that for scalar, EM, and acoustic waves, there is little structure in $v_E$ for low $f_s$ and as $f_s$ increases pronounced dips develop. In Fig. 1, we present the results of $v_E$ as a function of frequency for the acoustic case for three concentrations, confirming that the behavior observed by Schriemer et al. [1] is reproduced.

In addition, the new effective medium theory gives the correct long-wavelength limit for the effective dielectric constant $\epsilon_e$ for both the scalar and EM cases as well as the correct long-wavelength limit for the effective bulk modulus for the acoustic case.

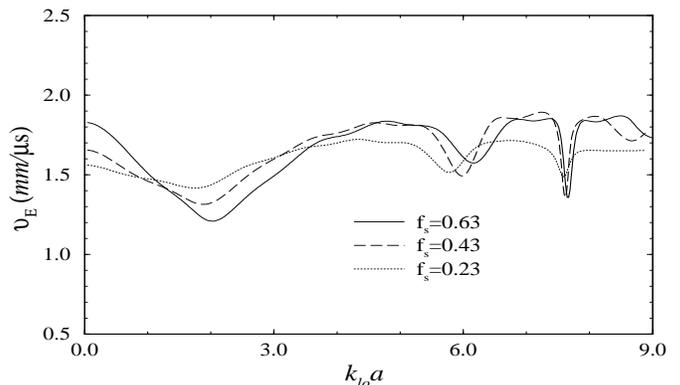

FIG. 1. Energy transport velocity, $v_E$, as a function of the dimensionless frequency $k_o a = \omega a/c_o$ for acoustic waves, propagating in a suspension of glass beats (of radius $a$) in water. $f_s$ is the volume fraction of the glass and $c_o$ the sound velocity in the water.

Sheng and his collaborators [1,2] have calculated the effective medium by finding the maximum of the imaginary part of the effective Green's function, the so-called spectral function. Finding the maximum of the spectral function, although physically very reasonable, is only an "Ansatz" just like the new effective medium theory [3,4] and has not been derived from a microscopic theory.

In conclusion, we feel there is no unique self-consistent way in determining the effective medium. Both the new effective medium theory [3,4] and the GCPA [2] give reasonable and well converged results for all wavelengths and scattering configurations. However, the most interesting open problem is to establish a detailed microscopic basis for these effective medium theories.

C. M. Soukoulis[1], K. Busch,[2] M. Kafesaki,[3] and E. N. Economou[3]

[1]Ames Laboratory and Dept. of Physics and Astronomy, Iowa State University, 50011, U.S.A.

[2]Department of Physics, University of Toronto, Toronto, Ontario, Canada M5S LA7.

[3]Research Center of Crete, FORTH and Dept. of Physics, 71110 Heraklion, Crete.

[1] H.P. Schriemer et al., Phys. Rev. Lett. **79**, 3166 (1997).
[2] X.D. Jing, P. Sheng and M. Y. Zhou, Phys.Rev. Lett. **66**, 1240 (1991); Phys. Rev. A **46**, 6513 (1992); Physica A **207** 37 (1994).
[3] K. Busch and C. M. Soukoulis, Phys. Rev. Lett. **75**, 3442 (1995); Phys. Rev. B **54**, 893 (1996); A. Kirchner, K. Busch, and C. M. Soukoulis, Phys. Rev. B **57**, 277 (1998).
[4] M. Kafesaki and E. N. Economou, Europhys. Lett. **37**, 7 (1997).
[5] E. N. Economou *Green's Functions in Quantum Physics*, Springer-Verlag, Berlin, 1983; C. M. Soukoulis, S. Datta, and E. N. Economou, Phys. Rev. B **49**, 3800 (1994).

1